\documentclass[epsfig,psfig,aps,twocolumn,prb,showpacs]{revtex4-1}
\usepackage[colorlinks=true,citecolor=red,urlcolor=blue]{hyperref}
\usepackage{amsfonts}
\usepackage{graphicx}
\usepackage{epsfig}
\begin{document}
\title
{
Kohn anomaly of optical zone boundary phonons in uniaxial strained graphene: role of the electronic band structure \\ 
}
\author
{
Sonia Haddad$^{\ast}$ and Lassaad Mandhour
} 
\affiliation{
Laboratoire de Physique de la Mati\`ere Condens\'ee, D\'epartement de Physique,
Facult\'e des Sciences de Tunis, Universit\'e Tunis El Manar, Campus Universitaire 1060 Tunis, Tunisia
}
%
%

\begin{abstract}
One of the unique properties of graphene is its extremely high mechanical strength. Several studies have shown that the mechanical failure of graphene sheet under a tensile strain is 
due to the enhancement of the Kohn anomaly of the zone boundary transverse optical phonon modes. In this work, we derive an analytical expression of the Kohn anomaly parameter
$\alpha_{\vec{K}}$ of these phonons in graphene deformed by a uniaxial strain along the armchair direction. We show that, the tilt of Dirac cones, induced by the strain,
contributes to the enhancement of the Kohn anomaly under a tensile deformation and gives rise to a dominant contribution of the so-called {\it outer} intervalley mediated phonon processes. 
Moreover, the Kohn anomaly is found to be anisotropic with respect to the phonon wave vectors around the K point. This anisotropy may be at the origin of 
the light polarization dependence of the Raman 2D band of the strained graphene.
Our results uncover, not only, the role of the Kohn anomaly in the anisotropic mechanical failure of the graphene sheet, under strains applied along the armchair 
and zigzag directions, but shed also light on the doping induced strengthening of strained graphene.
\end{abstract}

\pacs{73.22.Pr,63.22.Rc,78.67.Wj}
\maketitle

\section{Introduction}
Despite the unique features of graphene, several drawbacks have to be overcome to integrate this material in optoelectronic devices. 
In particular, the lack of a bandgap is a problem standing in the way of using graphene in electronics \cite{revcastro,gap}. 
Moreover, it has been proven that it is not possible to
obtain a high temperature intrinsic superconductivity in this material regarding the weak electron-phonon coupling responsible of the 
superconducting state \cite{SC-Rev}. \

In the last few years, strain engineering has emerged as a powerful tool to control the optical and electronic properties of 2D materials 
\cite{strainRev1,strainRev2,strainRevGinea,Ago,Guinea12}.
Strained graphene has been, recently, a hot topic of interest since it is expected to open the way to new applications for flexible 
electronic devices where the graphene sheet is manipulated as an origami paper \cite{origami,straingr}.\

Deformed graphene under strain may also offer new physical insights, as the generation of exotic electronic states under giant 
pseudomagnetic field \cite{Crommie} and a relatively high temperature superconductivity at $T_c \sim 30$ K \cite{Si13,Uchoa}.
Although the vibrational spectrum of graphene is significantly changed under strain, no bandgap has been induced in the electronic dispersion up to the critical
strain amplitude $\epsilon \sim 20\%$ \cite{strain-value} before sample cracking \cite{Castrogap,gap-num,Guineagap}. 
However, it is found that under uniaxial strain, the Dirac points shift from the high symmetry points $K$ and $K^{\prime}$
located at the corner of the hexagonal Brillouin zone\cite{revcastro,Castrogap,Li14}.\

The signature of the strain on the electronic and vibrational properties of graphene could be probed by Raman spectroscopy which is found to be sensitive
to the strain \cite{Basko,Huang,Ni,Ninano,Mohiuddin,frank,frank11,Lee2012,Huang13,Son,Popov,Thomsen,Mohr,Narula,Popov}. In particular, the G peak, originating from the doubly degenerate $E_{2g}$ center zone phonons, splits into two peaks whose intensities
are strongly dependent on the incident light polarization \cite{Mohiuddin}. This dependence is found to be the fingerprint of the strain modified electronic
dispersion, which affects the Raman G band through the electron-phonon interaction \cite{Assili}.\

Due to its higher strain induced frequency shift, the 2D Raman peak is commonly used to determine the strength and the direction of the applied strain 
\cite{Huang,Ninano,frank11,Son,Mohr,Neumann,Wang17}.
This peak is due to the double resonant intervalley process involving transverse optical boundary phonons with wave vector $\vec{K}$ \cite{Basko}.\newline
The characteristic features of the 2D band, under strain, are found to be substantially dependent on both the electronic structure and 
the dispersion of the inplane transverse optical phonon (iTO) mode at $K$ point \cite{Son,Popov,Thomsen,Mohr}. This dispersion is marked 
by a remarkable Kohn anomaly (KA) revealed by a pronounced kink which reflects a strong electron-phonon coupling (EPC)
\cite{Reich,Thomsen00,Basko08,Saito,Thomsen04,venezeula}. \

The KA occurs in metals due to the electron screening of the ionic potential \cite{Kohn}. 
This anomaly appears in the phonon branch as a sudden dip at a phonon wave vector $\vec{q}$
connecting two electronic states $\vec{k}$ and $\vec{k}^{\prime}$ on the Fermi surface satisfying $\vec{k}^{\prime}=\vec{k}+\vec{q}$.\

In graphene, the KA takes place at $\Gamma$ ($\vec{q}=\vec{0}$) and at $K$ point ($\vec{q}=\vec{K}$) since 
the Fermi surface reduced to the two points $K$ and $K^{\prime}$ \cite{Mauri04}.\

KA in non deformed graphene has been studied under close scrutiny \cite{Mauri04,Mauri06,Mauri08,Sasaki,KA}
since it measures the electron-phonon coupling which is a key parameter to understand several properties of graphene, such as the electronic transport, 
the stability of the superconducting state and the Raman spectra.\

However, a few studies can be found, in the literature, on the behavior of the KA in strained graphene. The role of KA has been shown to be crucial
for the mechanical failure process of pure graphene \cite{Marianetti,Si}. 
Si {\it et al.} \cite{Si} reported, based on first principles calculations, that the
strain induced enhancement of the KA in graphene could be counterbalanced by doping.
Recently, Cifuentes-Quintal {\it et al.} \cite{Bohnen16} showed that, besides the pronounced KA of the transverse optical phonon modes 
a new KA emerges, under a uniaxial strain, in the longitudinal acoustic phonon branch around the $K$ point.\

The outcomes of the studies, dealing with the behavior of the KA in strained graphene, pointed out several open questions. 
In particular, it is not understood why the doping induced weakening of the KA is much more pronounced in the strained graphene than
that in the unstrained lattice. On the other hand, the anisotropic failure mechanism of graphene sheet under zigzag and armchair tensile deformations 
is not clear.
Moreover, the behavior of the 2D Raman band under strain is not completely unveiled. Besides the hot debate on the type of the optical phonons
responsible of this bands, the origin of its light polarization dependence is still not fully understood \cite{Mohr,Narula}. \

Based on an analytical analysis of the KA mechanism in strained graphene, we try, in this paper, to provide some answers 
to the above mentioned puzzling points.\

We consider a honeycomb lattice under uniaxial strain applied along the armchair edges ($y$ axis). 
We neglect, hereafter, the strain component $\epsilon_{xx}$, along the $x$ axis perpendicular to the strain direction, regarding the 
small value of the Poisson ratio of graphene ($\nu=0.165$). This ratio relates the strain components as 
$\epsilon_{xx}=-\nu \epsilon_{yy}$ \cite{Poisson,Assili}.
Moreover, we do not consider the strain effect on the phonon band in order to highlight the signature of the electronic dispersion.
Such approximation was also used in Ref.\onlinecite{Mohr} to study the strain induced splitting of the 2D Raman band.\

The main results of this work could be summarized in the following points :
(i) The strain modified electronic dispersion affects substantially the KA. 
In particular, the tilt of Dirac cones is found to enhance the KA under a tensile deformation and to further the
so-called {\it outer} phonon mediated intervalley electronic transitions.
(ii) The KA shows an anisotropic behavior as a function of the phonon wave vector around the zone boundary K point. This anisotropy 
contributes to the light polarization dependence of the 2D Raman peak in strained graphene. 
(iii) The weakening of the KA with electronic doping is found to be more pronounced in strained graphene than in unstrained lattice.
(iv) The KA behavior gives rise to a large critical tensile deformation along the zigzag (ZZ) direction compared to the armchair axis, 
in agreement with the numerical calculations \cite{Liu,Zhao,Gao,PhysicaB,Polymer,John}.\

The paper is organized as follows: In section II, we derive the EPC expression for the graphene inplane TO phonon mode.
The behavior of the KA of these phonons is discussed in section III.  Sec. IV is devoted to the concluding remarks.

\section{Electron-phonon coupling: effective mass approach }

\subsection{Transverse optical phonon mode: KA slope}

We focus on the highest optical phonon branch at $K$ point corresponding to the $A^{\prime}_1$ mode showing a KA at a frequency $\omega_K$=161 meV \cite{Mauri04}.\

In graphene, the Fermi surface reduces to the points $K$ and $K^{\prime}$ and the density of states $N_F$, at the Fermi energy, is zero. 
As a consequence, the usual EPC coupling constant $\lambda _{\vec{q}}$, depending on $1/N_F$, is not well defined \cite{Mauri04}.
Theretofore, the EPC in graphene is, rather, characterized by the ratio $2\langle g^2_{\vec{q}}\rangle_F/{\hbar\omega_{\vec{q}}}$ 
where $2\langle g^2_{\vec{q}}\rangle_F$ is the average over the Fermi surface of $\arrowvert g_{\vec{k}+\vec{q},\lambda;\vec{k} \lambda^{\prime}}\arrowvert ^2$, 
and $g_{\vec{k}+\vec{q},\lambda;\vec{k} \lambda^{\prime}}$ is the coupling matrix element of the phonon with a wave vector $\vec{q}$ 
and electron in the state $\vec{k}$ within the band $\lambda$, which scatter to $\lambda^{\prime}$ band at the state 
$\vec{k}^{\prime}=\vec{k}+\vec{q}$ \cite{Mauri04}.\

In non-deformed graphene, the largest value of EPC is found for the $A^{\prime}_1$ mode for which 
$2\langle g^2_{K}\rangle_F/{\hbar\omega_{\vec{K}}}$=1.23 eV \cite{Mauri04}. 
This mode exhibits a KA described by a non zero slope $\alpha_{\vec{K}}$ of the phonon dispersion which can be written
around, the $K$ point, as $\omega_{\vec{K}+\vec{q}}= \alpha_{\vec{K}} \arrowvert \vec{q}\arrowvert + \hbar \omega_{\vec{K}} +  \mathcal{O}(q^2)$\cite{Mauri04}.
$\alpha_{\vec{K}}$ is related to the EPC by \cite{Mauri04}:
\begin{eqnarray}
\alpha_{\vec{K}}=\hbar \lim_{\vec{q}\rightarrow \vec{0}}
\frac{\omega_{\vec{K}+\vec{q}}-\omega_{\vec{K}}}{q}
=\hbar  \lim_{\vec{q}\rightarrow \vec{0}}
\frac{\tilde{D}_{\vec{K}+\vec{q}}-\tilde{D}_{\vec{K}}}{2\omega_{\vec{K}}M q}
\label{alpha}
\end{eqnarray}
where $q=\arrowvert \vec{q}\arrowvert$, $M$ is the carbon atomic mass and $\tilde{D}_{\vec{q}}$ is the non analytical component of the dynamical matrix \cite{Mauri04}
given by 
\begin{eqnarray}
\tilde{D}_{\vec{q}}=\frac{8M\omega_{\vec{K}}}{\hbar} \frac{S}{\left(2\pi^2\right)} \int d\vec{k}
\frac{\arrowvert g_{2\vec{K}+\vec{k}+\vec{q},\pi^{\ast} ;\vec{K}+\vec{k},\pi}\arrowvert ^2}
{\varepsilon_{\vec{K}+\vec{k},\pi}-\varepsilon_{2\vec{K}+\vec{k}+\vec{q},\pi^{\ast}}}
\label{tildeD}
\end{eqnarray}
where the transition of an electron from the occupied band ($\pi$) of the $K$ valley to the empty band ($\pi^{\ast}$) of the valley 
$K^{\prime}$ ($\vec{K^{\prime}}=2\vec{K}$) is considered.\\

In non-deformed graphene, the matrix element is of the form:
\begin{eqnarray}
\arrowvert g_{2\vec{K}+\vec{k}+\vec{q},\pi^{\ast} ;\vec{K}+\vec{k},\pi}\arrowvert ^2=\langle g^2_K\rangle_F \left( 1+\cos \theta\right)
\end{eqnarray}
where $\theta$ is the angle between $\vec{k}$ and $\vec{k}+\vec{q}$ \cite{Mauri04}.\

Considering the linear electronic dispersion around the Dirac points, $\alpha_{\vec{K}}$ becomes\cite{Mauri04} :
\begin{eqnarray}
\alpha_{\vec{K}}=\frac{8 \langle g^2_K\rangle_F}{\hbar v_F} 
\frac{S}{\left(2\pi^2\right)}\lim_{\vec{q}\rightarrow \vec{0}} 
\int_{k<k_m} d\vec{k}\left[ \frac 1{k}-
\frac{1+\cos \theta}{k+\arrowvert\vec{k}+\vec{q}\arrowvert}\right]
\label{alpha-integ} 
\end{eqnarray}
where $k_m$ is a cutoff corresponding to the limit of the linear dispersion of the electronic band.\
The numerical integration of the above expression gives $\alpha_{\vec{K}}\sim 253$ cm$^{-1}.\AA$ \cite{Mauri04,Mauri08}.\

In the following, we derive the expression of the KA slope $\alpha_{\vec{K}}$ in strained graphene. We first start by determining the EPC matrix element
$g_{2\vec{K}+\vec{k}+\vec{q},\pi^{\ast} ;\vec{K}+\vec{k},\pi}$.\\

\subsection{EPC in strained graphene}
\subsubsection{Electronic Hamiltonian}
\begin{figure}[hpbt]
\includegraphics[width=0.5\columnwidth]{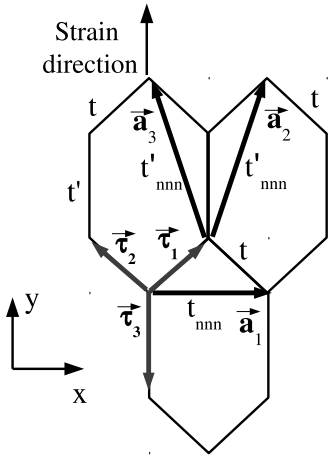}
\caption
{Deformed honeycomb lattice along the armchair $y$ axis.
($\vec{a}_1,\vec{a}_2$) is the lattice basis. 
The hopping parameters to the first (second) neighbors $t$ and $t^{\prime}$ ($t_{nnn}$ and $t_{nnn}^{\prime}$) are different due to the deformation. 
Vectors connecting first (second) neighboring atoms are denoted $\vec{\tau}_l$ ($\vec{a}_l$).}
\label{AClattice}
\end{figure}

We assume that the uniaxial strain is along the armchair (AC) direction, denoted $y$ axis, 
which results in a quinoid lattice \cite{mark2008} as shown in figure \ref{AClattice}.

The distance between nearest neighbor atoms, along the strain axis, changes from $a$ to $a^{\prime}=a+\delta a=a(1+\epsilon)$ where 
$\epsilon=\frac{\delta a} a$ is the lattice deformation or the strain amplitude. For a compressive (tensile) deformation $\epsilon$ is negative 
(positive).
The lattice basis is given by $(\vec{a}_1,\vec{a}_2)$ where
\begin{equation}
\vec{a}_1=\sqrt{3}a\vec{e}_x,\,
\vec{a}_2=\frac{\sqrt{3}}2a\vec{e}_x+a\left(\frac32 +\epsilon\right)\vec{e}_y
\end{equation}
The vectors joining the first neighbor atoms are:
\begin{eqnarray}
\vec{\tau}_1&=&\frac a 2 \left( \sqrt{3}\vec{e}_x+\vec{e}_y\right),\;
 \vec{\tau}_2=\frac a 2 \left( -\sqrt{3}\vec{e}_x+\vec{e}_y\right),\;\nonumber\\
\vec{\tau}_3&=&-a(1+\epsilon)\vec{e}_y.
\label{tau}
\end{eqnarray}
The hopping integral along $\vec{\tau}_3$ direction is modified by the strain from $t$ to $t^{\prime}=t+\frac{\partial t}{\partial a}\delta a $.\
The hopping terms to the second neighboring atoms change also compared to their values in undeformed graphene as 
\begin{eqnarray}
 t^{\prime}_{nnn}=t_{nnn}+\frac{\partial t_{nnn}}{\partial a}\delta a 
\end{eqnarray}
Assuming the Harrison law \cite{MarkRev}, $\frac{\partial t}{\partial a}=-\frac {2t}a$, then $t^{\prime}$ reduces to 
\begin{eqnarray}
t^{\prime}=t(1-2\epsilon)
\label{t}
\end{eqnarray}

It is worth to note that, beyond the linear elastic regime, the Harrison law is not accurate to deal with the strain induced changes of the hopping integrals \cite{EdouardoP}.
For a more accurate approach, it has been proposed to consider the hopping parameters deduced from Density Functional Theory (DFT) calculations \cite{Edouardo15}.

The quinoid lattice could be described by the so-called minimal form of the generalized 2D Weyl Hamiltonian \cite{MarkRev,suzumura2}
given by:
\begin{eqnarray}
 H_{\xi}(\vec{k})=\xi\left( \vec{w}_0.\vec{k}\sigma^0+w_xk_x\sigma^x\right)+w_yk_y\sigma^y
\label{Helec}
\end{eqnarray}
where $\sigma^0={1\!\!1}$, $\sigma^x$ and $\sigma^y$ are the 2x2 Pauli matrices, $\xi$ is the valley index, 
$w_x$ and $w_y$ characterize the anisotropy of the Dirac cones whereas $\vec{w}_0=(w_{0x},w_{0y}=0)$ is the tilt term.\
These parameters could be expressed, for small deformation amplitude ($|\epsilon|\ll1$), as \cite{MarkRev}:
\begin{eqnarray}
w_x=\frac 32 at (1+\frac 23 \epsilon),\,w_y=\frac 32 at (1-\frac 43 \epsilon), \,w_{0x}=0.6\,\epsilon\, w_x. \nonumber\\
\label{w}
\end{eqnarray}
The eigenenergies of the Weyl Hamiltonian, given by Eq.\ref{Helec}, are \cite{MarkRev}:
\begin{eqnarray}
 \varepsilon_{\lambda}(\vec{k})=\xi\vec{w}_0.\vec{k}+\lambda \sqrt{w_x^2k_x^2+w_y^2k_y^2}
\label{dispers}
\end{eqnarray}
The corresponding eigenfunctions are of the form:
\begin{eqnarray}
 \arrowvert\lambda,\vec{k}\rangle_D=\frac 1{\sqrt{2S^\prime}}
e^{i\vec{k}.\vec{r}} \left(
 \begin{array}{c}
 \lambda\\
 e^{i\Phi(\vec{k})}
 \end{array}
 \right) \\ \nonumber
 \arrowvert\lambda,\vec{k}\rangle_{D^{\prime}}=\frac 1{\sqrt{2S^\prime}}
 e^{i\vec{k}.\vec{r}}\left(
 \begin{array}{c}
 e^{i\Phi(\vec{k})}\\
 \lambda
 \end{array}
 \right)
\label{estate}
\end{eqnarray}
where $\Phi(\vec{k})$ is given by:
\begin{eqnarray}
\tan \Phi(\vec{k})=\frac {w_y k_y}{w_x k_x}
\label{Phik}
\end{eqnarray}

Under the strain, the Dirac cones are no more at the high symmetry points $K$ and $K^{\prime}$ but move according to \cite{MarkRev}:
\begin{eqnarray}
k^D_y=0,\; \; k^D_x=\xi \frac 2{\sqrt{3}a}\arccos\left(-\frac {t^{\prime}}{2t}\right).
\end{eqnarray}

\subsubsection{Electron-phonon coupling}
We consider the effective mass approach, so-called $\vec{k}.\vec{p}$, method to derive the Hamiltonian describing 
the interaction between the electrons and the zone boundary transverse optical phonons in uniaxial strained graphene.
This Hamiltonian could be obtained considering the effect of the lattice displacements on the hopping integrals in 
the undeformed electronic Hamiltonian. This approach was used by Ando \cite{Ando} in the case of undeformed graphene to obtain the EPC in the case
of the highest frequency zone boundary optical phonon mode. The authors applied the $\vec{k}.\vec{p}$ method for the electronic states around the Dirac valleys.
Within this method, the electronic wave function could be written as\cite{ando2006,Macucci}:
\begin{eqnarray}
 \psi(\vec{r})=\sum_{\vec{R}_A}\psi_A(\vec{R}_A) \varphi(\vec{r}-\vec{R}_A)+
\sum_{\vec{R}_B}\psi_B(\vec{R}_B) \varphi(\vec{r}-\vec{R}_B)\nonumber\\
\end{eqnarray}
where $\varphi(\vec{r}-\vec{R}_A)$ and $\varphi(\vec{r}-\vec{R}_B)$ are the atomic orbitals centered on atoms A and B respectively. 
The coefficients $\psi_A(\vec{R}_A)$ and $ \psi_B(\vec{R}_B)$ are expressed in terms of slowly varying envelope functions 
$F_A^D, \,F_A^{D^{\prime}},\, F_B^D$ and $F_B^{D^{\prime}}$ \cite{Ando}:
\begin{eqnarray}
\psi_A(\vec{R}_A)=\mathrm{e}^{i\vec{k}^D.\vec{R}_A } F_A^D(\vec{R}_A)+
\mathrm{e}^{i\vec{k}^{D^{\prime}}.\vec{R}_A } F_A^{D^{\prime}}(\vec{R}_A)\nonumber\\
\psi_B(\vec{R}_B)=\mathrm{e}^{i\vec{k}^D.\vec{R}_B } F_B^D(\vec{R}_B)-
\mathrm{e}^{i\vec{k}^{D^{\prime}}.\vec{R}_B } F_B^{D^{\prime}}(\vec{R}_B)
\end{eqnarray}
The Weyl Hamiltonian given by Eq.\ref{Helec} could be derived within the $\vec{k}.\vec{p}$ method taking into account the hopping terms 
to the second nearest neighbors \cite{Assili}. The eigenproblem is written as:
\begin{eqnarray}
 \varepsilon \psi_A(\vec{R}_A)=-\sum_{l=1}^3 t^{(l)} \psi_B(\vec{R}_A-\vec{\tau}_l)
-\sum_{l=1}^6 t_{nnn}^{(l)} \psi_A(\vec{R}_A-\vec{a}_l)\nonumber\\
\varepsilon \psi_B(\vec{R}_B)=-\sum_{l=1}^3 t^{(l)} \psi_A(\vec{R}_B+\vec{\tau}_l)
-\sum_{l=1}^6 t_{nnn}^{(l)} \psi_B(\vec{R}_B-\vec{a}_l)\nonumber\\
\label{eigen}
\end{eqnarray}
where $\vec{a}_4=-\vec{a}_1$, $\vec{a}_5=-\vec{a}_2$, $\vec{a}_6=-\vec{a}_3$ and 
$t^{(l)}$ ($t_{nnn}^{(l)}$) the hopping integrals to the first (second) neighboring atoms along $ \vec{\tau}_l$ ($\vec{a}_l$) vectors.\

Considering the effect of the lattice vibrations on the hopping integrals, generates an extra term in Eq.\ref{eigen} expressing the correction to these hopping
integrals. This term gives rise to the EPC Hamiltonian \cite{Assili}. \

For simplicity we consider, as in Refs.\onlinecite{Mauri04,Ando} the effect of lattice displacements on the hopping integral $t^{(l)}$ between first neighboring
atoms located at $\vec{R}_A$ and $\vec{R}_A-\vec{\tau}_l$. Due to the lattice vibration, this integral changes to
\begin{eqnarray}
t^{(l)}+\frac{\partial t^{(l)}}{\partial d_l} \left[|\vec{\tau}_l+\vec{u}_A(\vec{R}_A)-\vec{u}_B(\vec{R}_A-\vec{\tau}_l)|-d_l\right] 
\label{tnn}
\end{eqnarray}
where $d_l=|\vec{\tau}_l|$, $d_1=d_2=a$ and $d_3=a(1+\epsilon)$. $\vec{u}_A(\vec{R}_A)$ and $\vec{u}_B(\vec{R}_B=\vec{R}_A-\vec{\tau}_l)$ are the lattice 
displacements. \

For the zone boundary optical phonon modes of a wave vector $\vec{q}$, taken around the Dirac points $D$ and $D^{\prime}$, 
these displacements could be written as \cite{Ando}
\begin{eqnarray}
\vec{u}_A(\vec{R}_A)=\mathrm{e}^{i\vec{k}^D.\vec{R}_A }\vec{u}_A^D(\vec{R}_A) +
\mathrm{e}^{i\vec{k}^{D^{\prime}}.\vec{R}_A } \vec{u}_A^{D^{\prime}}(\vec{R}_A)\nonumber\\
\vec{u}_B(\vec{R}_B)=\mathrm{e}^{i\vec{k}^D.\vec{R}_B }\vec{u}_B^D(\vec{R}_B) +
\mathrm{e}^{i\vec{k}^{D^{\prime}}.\vec{R}_B } \vec{u}_B^{D^{\prime}}(\vec{R}_B)
\label{u_ab}
\end{eqnarray}

The coefficients $\vec{u}_A^D(\vec{r})$, $\vec{u}_A^{D^{\prime}}(\vec{r})$, $\vec{u}_B^D(\vec{r})$ and $\vec{u}_B^{D^{\prime}}(\vec{r})$ are given by:
\begin{eqnarray}
\vec{u}_{A/B}^{D/D^{\prime}}(\vec{r})=\sum_{\mu,\vec{q}}&& \sqrt{\frac{\hbar}{2NM\omega_{\mu}(\vec{q})}} \vec{e}_{A/B,\mu}^{\,D/D^{\prime}}(\vec{q})\times\nonumber\\
&&\left(b_{D,\mu,\vec{q}}+b^{\dagger}_{D^{\prime},\mu,-\vec{q}}\right) \mathrm{e}^{i\vec{q}.\vec{r}}\nonumber\\
\end{eqnarray}
where $b^{\dagger}_{D/D^{\prime},\mu,\vec{q}}$ ($b_{D/D^{\prime},\mu,\vec{q}}$) is the creation (annihilation) operator of a phonon with a wave
vector $\vec{q}$ in the mode $\mu$ around the Dirac point $D$ or $D^{\prime}$. $\omega_{\mu}(\vec{q})$ is the corresponding frequency which will be taken, hereafter,
equal to the highest value of the frequency zone boundary optical modes $\omega_{\mu}(\vec{q})=\omega_K =161$ meV.\

In order to highlight the role of the electronic band structure on EPC, we will assume that the phonon dispersion is not affected by the strain.
Following the method described in the appendix, we obtain the following EPC Hamiltonian:
\begin{eqnarray}
H_{int}=-\frac{3t}{a}\beta_K\left(1+\frac 13 \epsilon\right)
\left(
\begin{array}{cc}
0  & \Delta_{D^{\prime}} \sigma_y\\
\Delta_{D} \sigma_y & 0
\end{array}
\right)
\end{eqnarray}
where $\sigma_y$ is the Pauli matrix, $\beta_K=-\frac{b}{t}\frac{\partial t}{\partial b}$ and $\Delta_{D/D^{\prime}}$ is given by:
\begin{eqnarray}
\Delta_{D/D^{\prime}}=\sqrt{\frac{\hbar}{2NM\omega_K}}
\sum_{\vec{q}}\left(b_{D/D^{\prime},\vec{q}}+b^{\dagger}_{D^{\prime}/D,-\vec{q}}\right)
e^{i\vec{q}.\vec{q}}\nonumber\\
\end{eqnarray}
For $\epsilon=0$, we recover the Hamiltonian derived by Suzuura and Ando \cite{Ando} for undeformed graphene.\

To discuss the KA strain dependence, one needs to determine the slope $\alpha_{\vec{K}}$ which depends on the EPC matrix element $g_{D^{\prime},\vec{k}^{\prime}=\vec{k}+\vec{q},\pi^{\ast};D,\vec{k},\pi}$ corresponding to the transition of an electron from the occupied band ($\pi$) of the valley $D$ to the empty band $\pi^{\ast}$ at the $D$ valley.
\subsubsection{KA slope}
Given the electronic states of Eq.\ref{estate}, the EPC matrix element takes the form
\begin{eqnarray}
\arrowvert g_{D^{\prime},\vec{k}^{\prime}=\vec{k}+\vec{q},\pi^{\ast};D,\vec{k},\pi}\arrowvert^2&=&
\frac 12 \left(\frac{3at}{b^2}\beta_K\left(1+\frac 13 \epsilon \right)\right)^2\nonumber\\
&\times&\frac{\hbar}{2NM\omega_K}\left[1-\cos\left(\Phi(\vec{k})-\Phi(\vec{k}^{\prime}\right)\right]
\nonumber\\
\end{eqnarray}
where $\tan\Phi(\vec{k}^{\prime})=\frac{w_y k^{\prime}_y}{w_x k^{\prime}_x}$ with $\vec{k}^{\prime}=\vec{k}+\vec{q}$.\

Considering the electronic dispersion of Eq.\ref{dispers}, the KA slope $\alpha_{\vec{K}}$, given by Eq.\ref{alpha}, becomes
\begin{widetext}
\begin{eqnarray}
\alpha_{\vec{K}}&=&4 \left(\frac{3at}{b^2}\beta_K\left(1+\frac 13 \epsilon \right)\right)^2
\frac{\hbar}{2NM\omega_K}
\frac{S^{\prime}}{(2\pi)^2} \nonumber \\
&\times&\lim_{\vec{q}\rightarrow \vec{0}} \frac 1 q
\int_{k<km} d^2k\left[
\frac 1{\sqrt{w_x^2k_x^2+w_y^2k_y^2 }}
-\frac{1+\cos\left(\Phi(\vec{k})+\Phi(\vec{k}^{\prime})\right)}
{\vec{w}_0.\vec{q}+\sqrt{w_x^2k_x^2+w_y^2k_y^2}+ \sqrt{w_x^2(k_x+q_x)^2+w_y^2(k_y+q_y)^2}}
\right]
\label{alpha_cal}
\end{eqnarray} 
\end{widetext}
Introducing the components $\tilde{k}_x=w_xk_x$, $\tilde{k}_y=w_yk_y$ and the dimensionless variable $\tilde{y}=\frac{\tilde{k}}{\sqrt{w_xw_y} q}$, 
the integral in Eq.\ref{alpha_cal} takes the form:
\begin{widetext}
\begin{eqnarray}
I(\varphi,\epsilon)= \frac 1 w\int_0^{\infty}d\tilde{y}\int_0^{2\pi}d\Phi
\left\{
\frac 1{\sqrt{\tilde{w}_x\tilde{w}_y}}
-\tilde{y}\left[1+\cos\left(\Phi-\Phi^{\prime}\right)\right] f^{-1}(\tilde{y},\Phi)\right\}
\label{integral}
\end{eqnarray}
\end{widetext}
where $N$ is the number of unit cells, $S^{\prime}$ is the total surface of the deformed lattice $\Phi=\Phi(\vec{k})$ (Eq.\ref{Phik}), 
$w=\frac 32 at$ and the function $f(\tilde{y},\Phi)$ is given by:
\begin{widetext}
\begin{eqnarray}
f(\tilde{y},\Phi)=
\tilde{w}_{0x}\cos\varphi+\tilde{y}\sqrt{\tilde{w}_x\tilde{w}_y}
+\left[\tilde{w}_x\tilde{w}_y\tilde{y}^2+\tilde{w_x}^2\cos^2\varphi
+\tilde{w}_y^2\sin^2\varphi+
2\tilde{y}\sqrt{\tilde{w}_x\tilde{w}_y}\left(\tilde{w}_x\cos\varphi\cos\Phi
+\tilde{w}_y\sin\varphi\sin\Phi\right)\right]^{\frac 12}\nonumber\\
\end{eqnarray} 
\end{widetext}

where $\varphi=(\vec{e}_x,\vec{q})$ is the phonon wave vector angle, 
$\tilde{w}_x$, $\tilde{w}_y$ and $\tilde{w}_{0x}$ are dimensionless parameters corresponding to the normalization, by $\frac 32 at$,
of respectively ${w_x}$, ${w_y}$ and ${w}_{0x}$. $\Phi^{\prime}$ is given by:
\begin{eqnarray}
\Phi^{\prime}=\arctan 
\frac{\tilde{y}\sqrt{\tilde{w_x}\tilde{w_y}}\sin\Phi+\tilde{w_y}\sin\varphi}
{\tilde{y}\sqrt{\tilde{w_x}\tilde{w_y}}\cos\Phi+\tilde{w_x}\cos\varphi}
\end{eqnarray}
Given the Harrison's law\cite{mark2008}, $\frac{b}{t}\frac{\partial t}{\partial b}=-\frac{2}{d_l^2}$, 
and expressing the surface $S^{\prime}$ of the deformed lattice as a function of the 
undeformed one $S^{\prime}=N\parallel \vec{a}_1\times\vec{a}_2\parallel\simeq S\left(1+\frac 2 3 \epsilon \right)$,
the prefactor in the expression of $\alpha_{\vec{K}}$ (Eq.\ref{alpha_cal}) becomes:
\begin{eqnarray}
C_K=\frac{\sqrt{3} a_0 t}{2\pi}\lambda_K \left(1+\frac 4 3 \epsilon \right)
\end{eqnarray}
where $a_0=\sqrt{3} a$ is the lattice parameter 
and $\lambda_K$ is a dimensionless coupling parameter given by \cite{Ando}
\begin{eqnarray}
\lambda_K= \frac{36\sqrt{3}}{\pi}\frac{\hbar^2}{2Ma_0^2} \frac 1{\hbar\omega_K}.
\end{eqnarray}
For $\hbar\omega_K =161.2$ meV, $\lambda_K=\simeq 3.5 10^{-3}$.\

In the undeformed lattice, the integral given by Eq.\ref{integral} reduces to $\pi^2/2$ and taking $t=2.68$eV and $a_0=2.46 \AA$, gives $\alpha_{\vec{K}}=253 \,\rm{cm}^{-1}\AA$
(Eq.\ref{alpha_cal}) in agreement with numerical calculations \cite{Mauri04,Mauri06}.\

In the following, we discuss the role of the electronic band structure on the strain dependence of the KA slope $\alpha_{\vec{K}}$.

\section{Results and discussion}

\begin{figure}[hpbt] 
\begin{center}
\includegraphics[width=1\columnwidth]{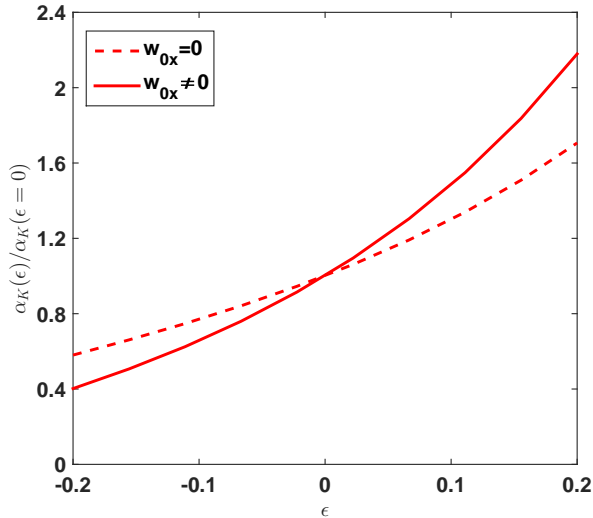}
\end{center}
\caption{KA slope $\alpha_{\vec{K}}$ as a function of strain. 
The data are normalized with respect to the value of $\alpha_{\vec{K}}$ for the unstrained graphene ($\epsilon=0$). 
The solid line corresponds to the deformed Dirac cones including anisotropy and tilt effects while the dashed one is calculated for the untilted cones ($w_{0x}=0$ in Eq.\ref{dispers}).
The phonon wave vector $_vec{q}$ is along the $x$ axis ($\varphi=0$).}
\label{KA-eps}
\end{figure}

Figure \ref{KA-eps} shows the normalized KA slope $\alpha_{\vec{K}}$ as a function of the strain. 
The normalization is taken with the respect to the slope value for unstrained lattice ($\epsilon=0$). 
The calculations are done for compressive ($\epsilon<0$) and tensile ($\epsilon>0$) deformations along the armchair $y$ axis. 
The phonon wave vector is taken along the $x$ axis ($\varphi=0$).

According to Fig.\ref{KA-eps}, the KA becomes more pronounced for tensile deformation. $\alpha_{\vec{K}}$ increases of about 
$20\%$ ($50\%$) for a strain of $5\%$ ($10\%$). However, a compressive deformation reduces the KA slope.
This behavior can be understood from the electronic band structure given by Eq.\ref{dispers}.\

Let us, first, disregard the tilt parameter $w_{0x}$. 
The shape of the Dirac cones depends on the electron velocities  $v_x= \frac{w_x}{\hbar}\sim v_F\left(1+\frac 23 \epsilon\right)$ and
$v_y= \frac{w_y}{\hbar}\sim v_F\left(1-\frac 43 \epsilon\right)$ (Eq.\ref{w}) where $v_F$ is the Fermi velocity in the unstrained lattice \cite{mark2008}.
Therefore, the electron velocity $v_y$ ($v_x$) along (perpendicular to) the strain direction decreases (increases) with the tensile deformation.\

We consider the phonon mediated intervalley electron scattering at a constant energy $E_L$ close to the Dirac points, as in the case of the double resonance Raman peak 2D,  
for which $E_L$ is the light excitation energy \cite{frank11,Narula,Popov}. 
For unstrained graphene, the momentum cutoff $k_m$ in equation \ref{alpha_cal} could be related to $E_L$ as $E_L\sim \hbar v_F k_m$.
Regarding the deformed Brillouin zone and the distorted Dirac cones, the intervalley processes become anisotropic.\

\begin{figure}[hpbt] 
\begin{center}
\includegraphics[width=1\columnwidth]{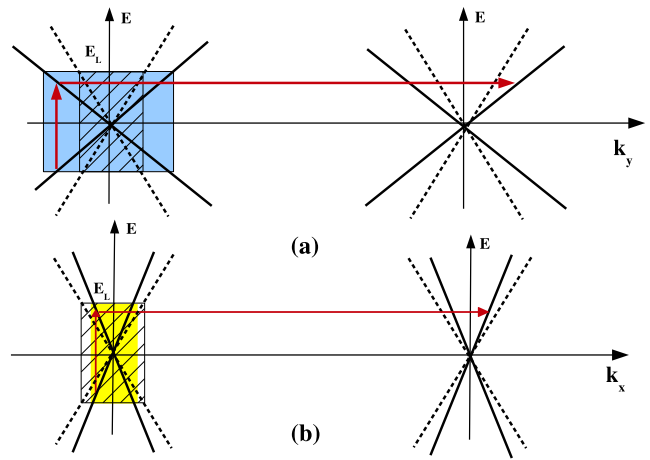}
\end{center}
\caption{Schematic representations of the intervalley phonon-mediated electronic transitions along the $k_y$ (a) and $k_x$ (b) axises at a given excitation energy $E_L$. 
The dashed (solid) lines represent the Dirac cones of the unstrained (strained) graphene. The Fermi velocity $v_y$ ($v_x$) along (perpendicular)
to the strain direction $(y^{\prime}y$) is reduced (enhanced) compared to isotropic case. 
This leads to more (less) electron-hole pairs contributing to the intervalley processes. 
The colored and dashed areas give the extension of the electronic states
contributing to the intervalley transitions around respectively anisotropic and isotropic Dirac cones.}
\label{kxy}
\end{figure}

Figure \ref{kxy} shows that, for a tensile strain, the number of electron-hole pairs involved in the intervalley scatterings is enhanced (reduced ) 
along the $y$ ($x$) axis compared to the undeformed lattice. 
Actually, this anisotropic scattering is schematically equivalent to excite an electron form the $\pi$ band, of the unstrained lattice, with a higher (lower) 
energy along the $k_y$ ($k_x$) axis.\

The tensile renormalization of $v_y$ is more pronounced than that of $v_x$, which means that globally,
the area of the electron wave vector $\vec{k}$ delimited by the equi-excitation energy contour $E_L$ is larger than that in unstrained graphene, 
which furthers the electron-phonon scatterings and enhances the KA slope $\alpha_{\vec{K}}$, as shown in Fig.\ref{KA-eps}.

For a compressive strain, the electron-phonon interaction is reduced since the deformation affects much more the processes along $y$ axis, for which the number of created electron-hole 
pairs is reduced regarding the strain induced enhancement of the electron velocity $v_y$.\\

\begin{figure}[hpbt] 
\begin{center}
\includegraphics[width=1\columnwidth]{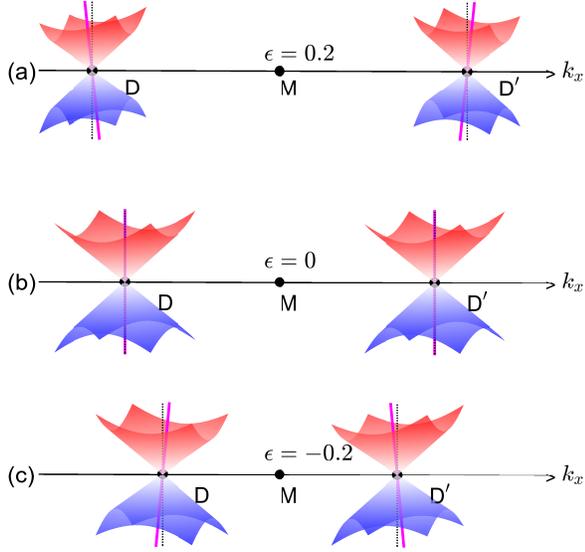}
\end{center}
\caption{Dirac cones in strained graphene ((a) and (c)) and in undeformed lattice (b). 
The colored axis give the tilt direction of the Dirac cone compared to the untilted case (dashed axis).
Under a tensile (compressive) deformation, the Dirac cones, along $DMD^{\prime}$ direction, are tilted towards the $\Gamma$  ($M$) point.}
\label{Figtilt} 
\end{figure}

Let us, now, discuss the role of the Dirac cone tilt.
According to Fig.\ref{KA-eps}, the KA for a tensile strain is more marked in the presence of the tilt while it is reduced for a compressive deformation.
To explain this behavior, we plot in figure \ref{Figtilt} the electronic dispersion, given by Eq.\ref{dispers}, along the $k_x$ axis, around the Dirac points 
in the direction $DMD^{\prime}$ for unstrained graphene and  deformed lattices under a tensile ($\epsilon=0.2$) and a compressive deformation ($\epsilon=-0.2$). 
The positions and the shape of the deformed Dirac cones 
are determined using the whole band structure of the quinoid lattice \cite{mark2008}. The Dirac cones move  towards each other (away) under a compressive (tensile) deformation \cite{MarkRev}

\begin{figure}[hpbt] 
\begin{center}
\includegraphics[width=0.8\columnwidth]{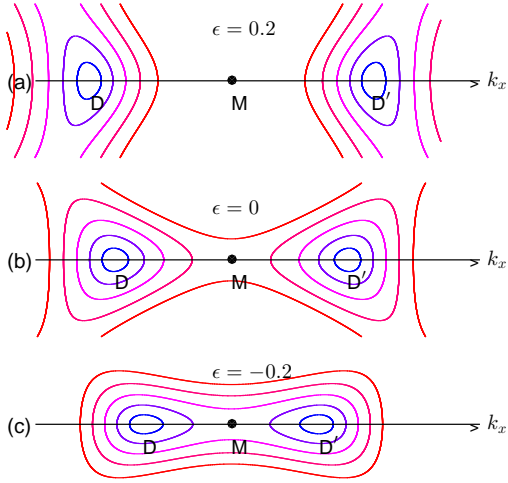}
\end{center}
\caption{Iso-energy contours along the $k_x$ axis for strained (a), (c) and undeformed (b) graphene.}
\label{Figcontour} 
\end{figure}

\begin{figure}[hpbt] 
\begin{center}
\includegraphics[width=1\columnwidth]{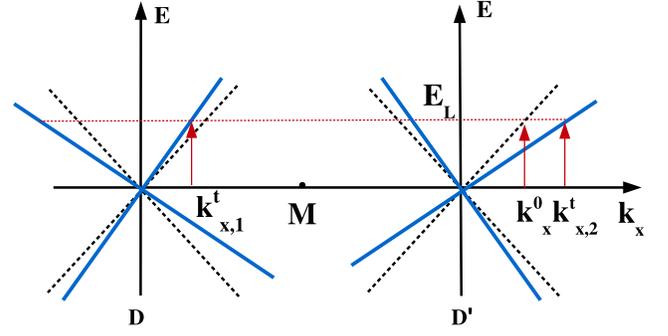}
\end{center}
\caption{Intervalley phonon-mediated electronic transitions along the $k_x$ direction under a tensile deformation and at the excitation energy $E_L$ indicated by the red dashed line.
The dashed cones represent the nontilted case with a Fermi velocity $v_x$ while the blue cones have also the same velocity $v_x$ but are tilted due to the term $w_{0x}$ in Eq.\ref{dispers}.
$k^0_{x}$, $k^t_{x,1}$ and $k^t_{x,2}$, indicated by the arrows, are the electronic states of, respectively, the unstrained and deformed lattices 
corresponding to the excitation energy $E_L$.}
\label{Figtilt-exp}
\end{figure}

The corresponding iso-energy contours are depicted in Fig.\ref{Figcontour} showing that, along the $DMD^{\prime}$ direction,
the curvature of the these contours is more affected for the outer (inner) electronic states under a tensile (compressive) deformation. 
By outer and inner states we refer respectively
to the states connected by a phonon wave vectors $q_{out}>k_D$ and $q_{in}<k_D$.\
For reasons of clarity, we, schematically, represented in figure \ref{Figtilt-exp} the band structure depicted in Fig.\ref{Figtilt} (a).\

For a tensile deformation, the outer states are in the tilting direction of the Dirac cones as shown in Fig.\ref{Figtilt-exp}, where
we set $k^0_{x}$ the electron wave vector corresponding the excitation energy $E_L$ in the non-tilted case, namely: $E_L=w_x k^0_{x}$.
We denote by $k^t_{x,1}$ and $k^t_{x,2}$ the wave vectors ascribed to the tilted cone in the D valley ($\xi=+1$ in Eq.\ref{dispers}) given by:
\begin{equation}
E_L=(w_{0x}+w_x)k^t_{x,1}, \,\,E_L=(w_x-w_{0x})k^t_{x,2}
\end{equation}

with $w_{0x}\sim 0.6\, \epsilon\, w_x$ (Eq.\ref{dispers}).\

The area of the equi-energy contour increases for the tilted cone since $k^t_{x,2}-k^0_{x}>k^t_{x,1}-k^0_{x}$, which gives rise to an enhanced number of electron-hole pairs. 
As a consequence, the electron-phonon interaction increases, which yields to the enhancement of the KA parameter $\alpha_{\vec{K}}$.
It is worth to note that the electronic states along the $k_y$ axis are not affected by the tilt of Dirac cones.\

According to Fig.\ref{Figtilt-exp}, this enhancement is due to the so-called {\it outer} intervalley processes involving phonons with wave vectors $q_{out}>k_D$.
The dominance of the inner or outer processes in the double resonance 2D Raman peak has been a hot topic of debate. Early experimental 
and numerical studies have argued that the outer phonons contribute mostly to the uniaxial strain induced splitting of the 2D mode \cite{Ferrari06,Thomsen00,Kurti}. 
This outcome was controverted by later findings based on numerical calculations and highlighting the dominance of the inner processes \cite{Huang,Mohr,Son}.  
Narula {\it et al.}\cite{Narula} have revoked the dominance of both type of phonons and showed, through a numerical study, that the dominant phonon wave vector 
is highly anisotropic and the distinction between inner and outer processes is irrelevant.\
It is worth to stress that the above mentioned numerical calculations take into account the strain induced change of the phonon dispersion which is not included in the present work.
Our result shows that, the tensile modified electronic dispersion, gives rise to a dominance of the outer phonons in the electron-phonon interaction process and 
this dominance is due to the tilt of Dirac cones. \

On the other hand, Narula {\it et al.}\cite{Narula} have found that the splitting of the 2D peak under a uniaxial tensile strain cannot originate only from the shift of the Dirac points. 
This result is in agreement with our work showing that the number of the electron-hole pairs involved in the electron-phonon interaction process is
independent of the Dirac cone position. This process depends basically on the shape of the equi-excitation energy contours governed by the parameters $w_x$ and $w_y$
and the tilt factor $\vec{w_0}$ (Eq.\ref{dispers}).\

The strain dependence of the KA depicted in figure \ref{KA-eps} could shed light on the anisotropic  mechanical properties of strained graphene. Ni {\it et al.} \cite{PhysicaB} have reported,
based on molecular dynamics models, that the AC tensile deformation causes the fracture of graphene sooner than a tensile applied along the zigzag (ZZ) edge. 
This anisotropic behavior was ascribed, by the authors, to different changes of the C-C bond angles. A larger critical strain along the ZZ axis was also reported 
in Refs.\onlinecite{Gao,Polymer}.\

In the following, we show that the KA could be responsible of the anisotropic failure mechanism of the graphene sheet.

\begin{figure}[hpbt] 
\begin{center}
\includegraphics[width=1\columnwidth]{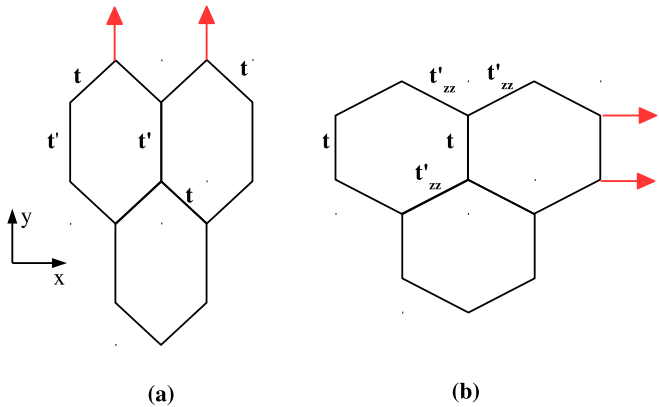}
\end{center}
\caption{Deformed honeycomb lattice under a tensile applied along the armchair direction (a) and zigzag one (b). The arrows indicate the direction of the tensile deformation.
$t$, $t^{\prime}$ and $t^{\prime}_{ZZ}$ are the hopping integrals between first neighboring atoms. }
\label{ZZ}
\end{figure}

In figure \ref{ZZ}, we depicted a schematic representations of the lattice deformed under AC and ZZ tensile where $t$ denote the hopping integral,
between first neighbors in unstrained system, and $t^{\prime}$ ($t^{\prime}_{ZZ}$) is the hopping integral under an AC (ZZ) tensile.

The lattice deformed under a ZZ tensile could be regarded as that obtained under a compressive AC strain with unstrained hopping $\tilde{t}=t^{\prime}_{ZZ}$ and a strain modified hopping integral
$\tilde{t}^{\prime}=t$ with $\tilde{t}^{\prime}=\tilde{t}(1+2|\epsilon|)$ (Eq.\ref{t}). The corresponding KA slope is that given by equation \ref{alpha_cal} 
by changing $t$ by $\tilde{t}$ and ${t}^{\prime}$ by $\tilde{t}^{\prime}$.\

As shown in figure \ref{KA-eps}, the KA is reduced, under a compressive strain, compared to the undeformed case ($\epsilon=0$). Moreover, changing $t$ by $\tilde{t}=t^{\prime}_{ZZ}<t$ 
in Eq.\ref{alpha_cal}, reduces the prefactor term and then weakens the KA. As a consequence, the KA is reduced for a ZZ tensile strain compared to the AC one. 
This result is consistent with the anisotropic frequency shifts of the TO phonon mode under strain along ZZ and AC directions obtained within first-principles 
calculations in Ref.\onlinecite{Bohnen16}
We then ascribe the relatively large critical strain along the ZZ edges to the hardening of the TO phonon modes at K point induced by the weakening of the KA. This interpretation
is different from that deduced from molecular dynamics calculations ascribing such behavior to the orientation of the C-C bonds 
with respect to the applied force direction \cite{Gao}.\

It is worth to note, that the lattice softening, resulting from the enhancement of the KA under tensile uniaxial strain, could be counterbalanced by charge doping as reported by Si {\it et al.}
\cite{Si}. The authors have found a peculiar behavior of the doping induced frequency shifts of the TO modes at $K$ point. In strained graphene, this shift is remarkably greater than 
that in unstrained lattice. The authors mentioned that the origin of this large difference is not clear within the framework of their first-principles calculations.
In the next, we give, based on schematic representations of the doped graphene band structure, a possible interpretation of this feature.\

\begin{figure}[hpbt] 
\begin{center}
\includegraphics[width=1\columnwidth]{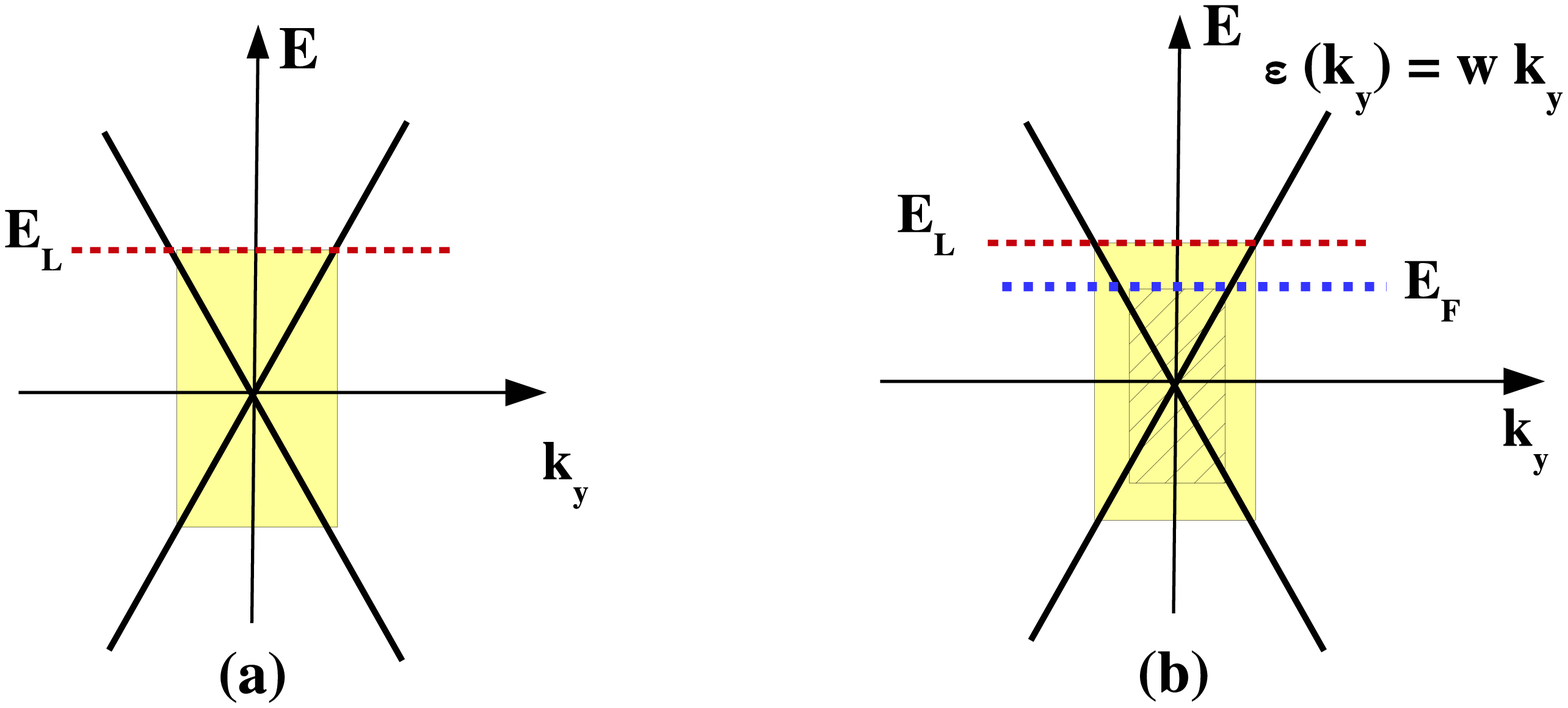}
\includegraphics[width=1\columnwidth]{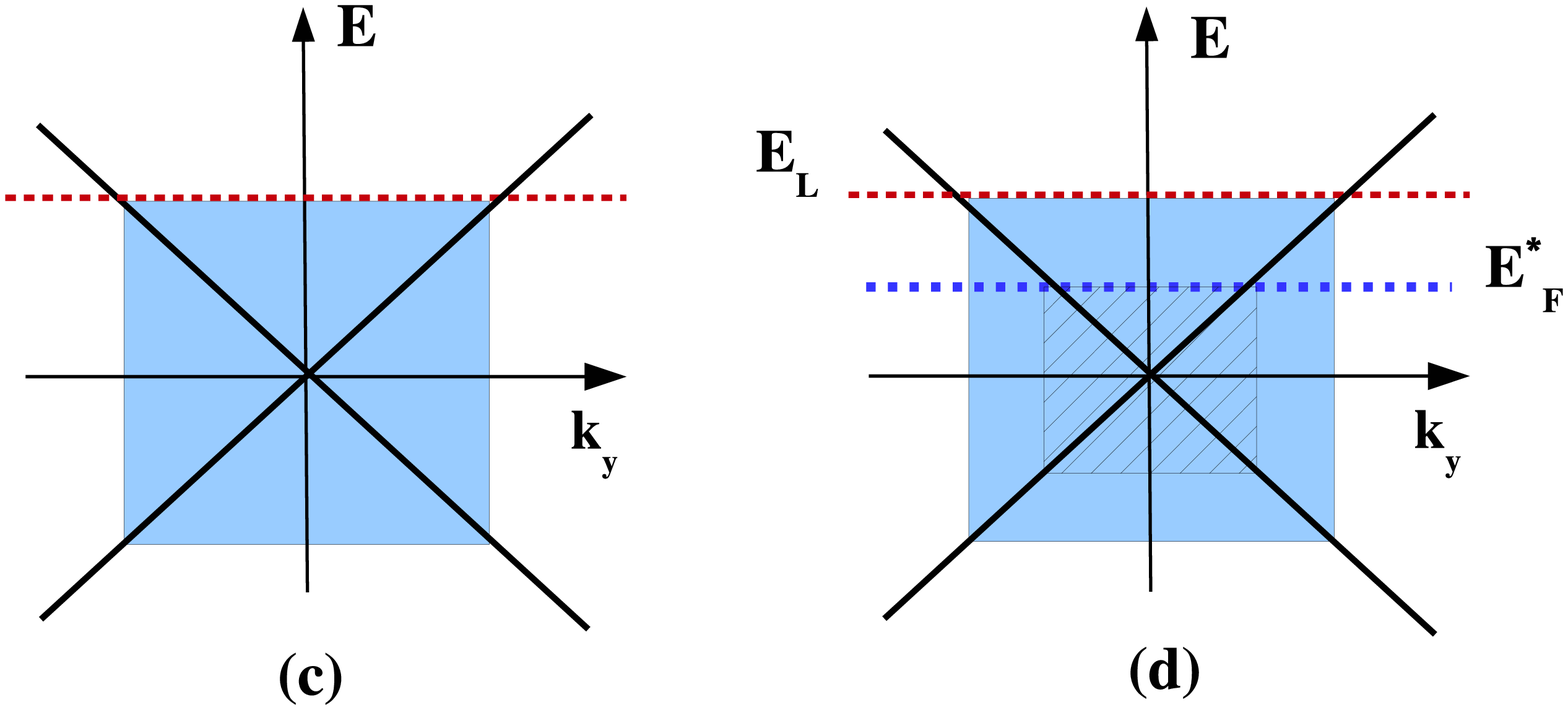}
\end{center}
\caption{Schematic representation of the electronic states along $k_y$ direction involved in the intervalley phonon-mediated 
processes for a given excitation energy $E_L$. (a) and (c) ((b) and ((d)) 
show to the undoped (doped) unstrained and strained lattices respectively. $E^{\ast}_F$ is the strain-renormalized Fermi energy. 
The dashed areas correspond to the electronic states blocked by the Pauli principle. }
\label{doped}
\end{figure}

Figure \ref{doped} (a) shows the electron-hole pairs involved in the intervalley phonon-mediated processes in unstrained graphene for a given excitation energy $E_L$ and at charge neutrality.
By electron doping at $E_F<E_L$ (Fig.\ref{doped} (b)), the number of the electron-hole pairs, contributing to the intervalley processes, is reduced due to Pauli principle, which explains the hardening of the phonon frequency by 
doping unstrained graphene \cite{Si}.

Under a uniaxial tensile, the Fermi energy is renormalized as \cite{mark2008}
\begin{eqnarray}
 E^{\ast}_F=E_F\left(1-\frac 13 \epsilon\right)
 \label{EF}
\end{eqnarray}

In figure \ref{doped}, we represented the electronic states involved in the intervalley processes along $k_y$ direction which give, as discussed above, a
dominant contribution to the KA under a tensile deformation.
The number of electronic states blocked by the Pauli principle in the strained lattice is greater than that for undeformed case. Indeed, these states are in the interval
$\Delta k_y=E^{\ast}_F/w_y\sim (1+ \epsilon)E_F/w$ while in the unstrained lattice, the locked states are within the interval
$\Delta k^{0}_y=E_F/w$ where $w=\frac 32 at$ ((Eq.\ref{dispers}). \

For $\epsilon>0$, $ \Delta k_y>\Delta k^{0}_y $. This leads to a larger number of blocked intervalley processes in graphene under uniaxial tensile strain which is consistent with the
result of Ref.\onlinecite{Si}. Moreover, the KA is expected to be weakened by doping regarding the enhancement of $ \Delta k_y$ with $E_F$ which is in agreement 
with Refs. \onlinecite{Si,Wang17}.\\

\begin{figure}[hpbt] 
\begin{center}
\includegraphics[width=1\columnwidth]{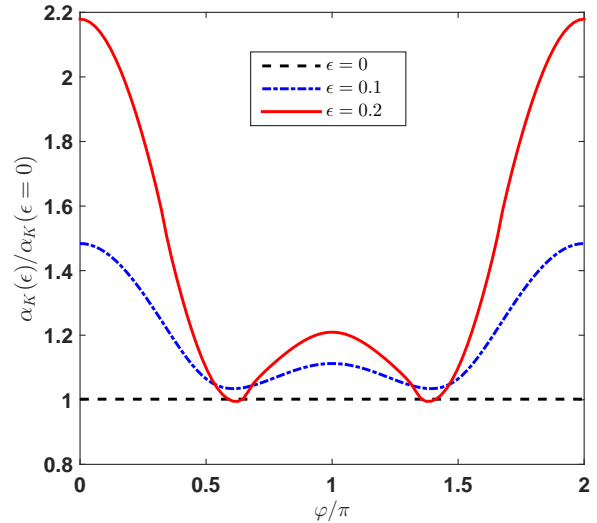}
\end{center}
\caption{Phonon angle dependence of the KA slope $\alpha_{\vec{K}}$ for different strain amplitudes. 
$\varphi$ is the angle between the phonon wave vector $\vec{q}$ and the $x$ axis perpendicular to the strain direction. 
The data are normalized with respect to the value of $\alpha_{\vec{K}}$ for $\varphi=0$ in the unstrained lattice ($\epsilon=0$).}
\label{FigKAphi}
\end{figure}

In figure \ref{FigKAphi} we represent the phonon angle dependence of the normalized KA parameter for different strain values. 
For the unstrained lattice, the KA is isotropic with respect to the phonon wave vector direction since the iso-energy contours are almost circular at low energy. 
As the strain amplitude increases, the KA becomes anisotropic with a dominant component for phonons with wave vector $\vec{q}$ along the $x$ axis which 
is perpendicular to the strain direction. \

This result is in agreement with the light polarization angle dependence of the 2D Raman band reported in literature.\
Several experimental and numerical studies \cite{Huang,frank11,Popov} have found that the 2D peak splits,
under strain, into two lines. Under AC strain, the intensity of the line associated to a parallel light polarization, 
with respect to the strain direction, is greater than that of the peak ascribed to the perpendicular polarization i.e. 
$I_{2D}(\theta=90^{\circ}) <I_{2D}(\theta=0^{\circ}) $, where $\theta$ is the angle of the light polarization with the respect to the strain direction \cite{Popov}.
Since the Raman intensity depends on $|\vec{q}\times \vec{E}|$ where $\vec{q}$ is the phonon wave vector and $\vec{E}$ is the light polarization \cite{Basko}, the 2D band is 
then expected to show a large intensity for $\vec{q}$ along the direction perpendicular to the strain axis. This behavior is in agreement with our result depicted in figure \ref{FigKAphi}
showing that the KA is enhanced for $\varphi=0$. \

Moreover, figure \ref{FigKAphi} shows that the KA should exhibit a minimum around the $y$ direction ($\varphi=\pi/2$). This behavior is due to 
the tilt of Dirac cones which, as discussed above, enhances the KA for electronic states along $k_x$ axis.
According to figure \ref{FigKAphi}, the KA is expected to have a relative maximum for phonons with $\varphi=\pi$, which correspond to the inner intervalley processes. The latter have, 
as we already shown, have a minor contribution, to the KA, compared to the outer processes.

\begin{figure}[hpbt] 
\begin{center}
\includegraphics[width=1\columnwidth]{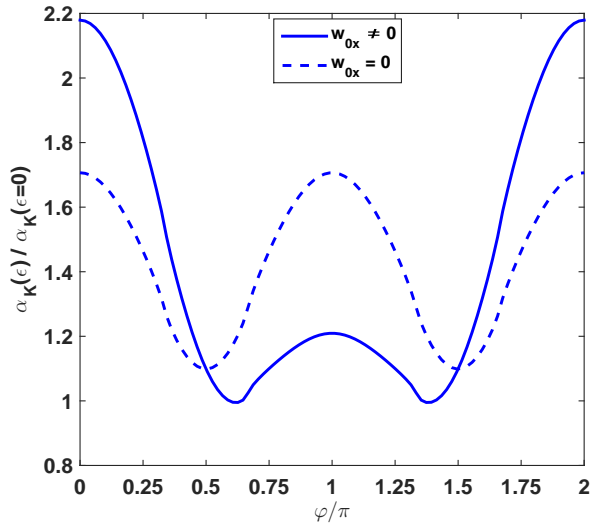}
\end{center}
\caption{Phonon angle dependence of the KA slope $\alpha_{\vec{K}}$ in the case of titled (solid line) and nontilted (dashed line) cones. 
The data are normalized with respect to the value of $\alpha_{\vec{K}}$ for $\varphi=0$ in the unstrained lattice.}
\label{FigKAphi02}
\end{figure}

Figure \ref{FigKAphi02} shows the KA slope as a function of the phonon angle for a tensile strain of $\epsilon=0.2$. T
he solid (dashed) curve corresponds to the case of tilted (non-tilted) 
Dirac cones. According to this figure, the anisotropic behavior of $\alpha_{\vec{K}}$ is due to the anisotropic Fermi velocities of the $v_x=w_x/\hbar$ and $v_y=w_y/\hbar$. 
However, the tilt parameter is responsible of the dominance of the outer intervalley phonon processes, corresponding to $\varphi=0$, for which $\alpha_{\vec{K}}$ reaches its maximum value.
The inner processes have a lower contribution associated to $\varphi=\pi$. \

As mentioned by Narula {\it et al.}, the notion of inner and outer processes is rather confusing since they can be mapped into
each other by the addition of a reciprocal lattice vector. The authors showed, based on numerical calculations, 
that the dominant phonon-mediated intervalley electronic transitions are neither inner nor outer but with a significant contribution of the inner processes.
To avoid any confusing nomenclature we conclude that the intervalley processes, connecting the most deformed parts of the electronic iso-energy contours, 
have the dominant contribution to the KA around the Dirac points.

\section{Conclusion}
In summary, we have presented an analytical study of the effect of the electronic dispersion relation on the KA of strained graphene. We found that, besides the shifts of the KA phonon
wave vector, the strain dependence of the slope parameter $\alpha_{\vec{K}}$ describing this anomaly is substantially dependent on the electronic band structure. 
In particular, the KA is found to be enhanced under tensile strain, by the tilt of Dirac cones. The latter furthers the so-called {\it outer} intervalley phonon processes.
We have, also, found that the strain dependence of the electronic band structure is at the origin of the strong doping induced reduction of the KA in graphene under a tensile 
deformation compared to the undeformed lattice.
Moreover, our results show that the KA is anisotropic with respect to the phonon wave vector which may give insights not only on light polarization dependence of Raman 2D band 
but also on the anisotropic mechanical failure of graphene under strain.

\section{Acknowledgment}
We thank M. E. Cifuentes-Quintal for stimulating discussions.
We are indebted to M. E. Cifuentes-Quintal, J.- N. Fuchs and Reza Asgari for a critical reading of the manuscript.
S. H. acknowledges the kind hospitality of ICTP (Trieste, Italy) where part of the work  was carried out. S. H. was supported by Simons-ICTP associate fellowship.

\appendix
\section{EPC Hamiltonian by $\vec{k}.\vec{p}$ method}

The $\vec{k}.\vec{p}$ method was used by Suzuura and Ando \cite{Ando} to obtain the effective Hamiltonian describing the interaction between electrons 
and the zone boundary optical phonons corresponding to the highest frequency mode, the so-called Kekul\'e mode.
This method was also used to determine the electron-phonon interaction Hamiltonian in the case of the optical center zone phonon modes of graphene in the absence of deformation \cite{ando2006} and under a uniaxial strain \cite{Assili}.\

Based on Ref.\onlinecite{Ando}, we derive the EPC matrix element $g_{D^{\prime},\vec{k}^{\prime}=\vec{k}+\vec{q},\pi^{\ast};D,\vec{k},\pi}$ 
corresponding to the transition of an electron from the occupied band ($\pi$) of the valley $D$ to the empty band $\pi^{\ast}$ at the $D^{\prime}$ valley 
in graphene, under uniaxial strain applied along the armchair direction.\

We start with the electronic eigenproblem given by Eq.\ref{eigen} where the functions $\psi_A(\vec{R}_A)$ and $\psi_B(\vec{R}_B)$ can be written in terms of the envelope functions $F_A^{D/D^{\prime}}(\vec{R}_A )$ and $F_B^{D/D^{\prime}}(\vec{R}_B )$ as :
\begin{eqnarray}
\psi_A(\vec{R}_A)=a^{\dagger}(\vec{R}_A)\Phi_A(\vec{R}_A)\nonumber\\
\psi_B(\vec{R}_B)=b^{\dagger}(\vec{R}_B)\Phi_B(\vec{R}_B)\nonumber\\
\end{eqnarray}
with
\begin{eqnarray}
a(\vec{R}_A)=\left(
\begin{array}{c}
\mathrm{e}^{-i\vec{k}^D.\vec{R}_A}\\
\mathrm{e}^{-i\vec{k}^{D^{\prime}}.\vec{R}_A}
\end{array}
\right)\quad
b(\vec{R}_B)=\left(
\begin{array}{c}
\mathrm{e}^{-i\vec{k}^D.\vec{R}_B}\\
-\mathrm{e}^{-i\vec{k}^{D^{\prime}}.\vec{R}_B}
\end{array}
\right)\nonumber\\
\Phi_A(\vec{R}_A)=\left(
\begin{array}{c}
F_A^D(\vec{R}_A)\\
F_A^{D^{\prime}}(\vec{R}_A)
\end{array}
\right)\quad
\Phi_B(\vec{R}_B)=\left(
\begin{array}{c}
F_B^D(\vec{R}_B)\\
F_B^{D^{\prime}}(\vec{R}_B)
\end{array}\right)\nonumber\\
\end{eqnarray}
As in Ref.\onlinecite{Ando}, we introduce the smoothing function $g(\vec{r})$ satisfying the following relations:
\begin{eqnarray}
\sum_{\vec{R}_A} g(\vec{r}-\vec{R}_A)=\sum_{\vec{R}_B} g(\vec{r}-\vec{R}_B)=1\nonumber\\
f(\vec{r})g(\vec{r}-\vec{R}_A)\simeq f(\vec{R})g(\vec{r}-\vec{R}).
\end{eqnarray}
where $f(\vec{r})$ is an envelope function\cite{ando2006}.
The left-hand side of Eq.\ref{eigen} can then be written, at $\vec{R}_A$ site, as:
\begin{widetext}
\begin{eqnarray}
\varepsilon\, a(\vec{R}_A)a^{\dagger}(\vec{R}_A)F_A(\vec{r})
&=&\varepsilon \sum_{\vec{R}_A}g(\vec{r}-\vec{R}_A)a(\vec{R}_A)a^{\dagger}(\vec{R}_A)F_A(\vec{r})
=-\sum_{l=1}^3 t^{(l)}\sum_{\vec{R}_A} g(\vec{r}-\vec{R}_A)a(\vec{R}_A)b^{\dagger}(\vec{R}_B)F_B(\vec{r}-\vec{\tau}_l)\nonumber\\
&-&\sum_{l=1}^6 t_{nnn}^{(l)}\sum_{\vec{R}_A} 
g(\vec{r}-\vec{R}_A)a(\vec{R}_A)a^{\dagger}(\vec{R}_A-\vec{a}_l)F_A(\vec{r}-\vec{a}_l)
\label{eigen2}
\end{eqnarray} 
\end{widetext}
For small strain amplitude, the following relations are satisfied:
\begin{widetext}
\begin{eqnarray}
&&\sum_{\vec{R}_A} g(\vec{r}-\vec{R}_A)\mathrm{e}^{i(\vec{k}^{D^{\prime}}-\vec{k}^{D}).\vec{R}_A}=
\sum_{\vec{R}_B}
g(\vec{r}-\vec{R}_B)\mathrm{e}^{i(\vec{k}^{D^{\prime}}-\vec{k}^{D}).\vec{R}_B}\simeq 0\nonumber\\
&&\sum_{\vec{R}_A} g(\vec{r}-\vec{R}_A)\mathrm{e}^{i\vec{k}^{D}.\vec{R}_A}\simeq 0
\end{eqnarray} 
\end{widetext}

Taking into account the lattice vibrations on the hopping integral to the first neighbor atoms, an extra term appears in the eigenproblem given by Eq.\ref{eigen}. This term is of the form:
\begin{widetext}
\begin{eqnarray}
H_{int}F_B(\vec{r})=\sum_l\sum_{\vec{R}_A}g(\vec{r}-\vec{R}_A)a(\vec{R}_A)
b(\vec{R}_A-\vec{\tau}_l)
\left(-\frac{\partial t^{(l)}}{\partial d_l}\right)\left(\frac{\vec{\tau}_l}{d_l}\right).
\left(\vec{u}_A(\vec{R}_A)-\vec{u}_B(\vec{R}_A-\vec{\tau}_l)\right)F_B(\vec{r})
\label{int}
\end{eqnarray} 
\end{widetext}
where $F_B(\vec{r})=\left(
\begin{array}{c}
F_B^D(\vec{r})\\
F_B^{D^{\prime}}(\vec{r})
\end{array}
\right)$.\

We consider the phonon modes around the Dirac points $D$ and $D^{\prime}$ with wave vector $\vec{k}^{D/D^{\prime}}+\vec{q}$ where $\|\vec{q}\|\ll \frac{2\pi}a$. We can then use the continuum limit and put in Eq.\ref{u_ab}:
$\vec{u}^{D/D^{\prime}}_A(\vec{R}_A)\simeq \vec{u}^{D/D^{\prime}}_A(\vec{r})$ and
$\vec{u}^{D/D^{\prime}}_B(\vec{R}_A-\vec{\tau}_l)\simeq \vec{u}^{D/D^{\prime}}_B(\vec{r})$.\
Equation \ref{int} can then be written as:
\begin{eqnarray}
H_{int}F^D_B(\vec{r})=h^{\prime AB}_{int}F^{D^{\prime}}_B(\vec{r})
\end{eqnarray}
where $h^{\prime AB}_{int}$ is given by:
\begin{widetext}
\begin{eqnarray}
h^{\prime AB}_{int}=\sum_l
\left(-\frac{\partial t^{(l)}}{\partial d_l}\right)\left(\frac{\vec{\tau}_l}{d_l}\right).
\left[\mathrm{e}^{-i\vec{k}^{D^{\prime}}.\vec{\tau}_l}\vec{u}^{D^{\prime}}_A(\vec{r})
-\mathrm{e}^{-2i\vec{k}^{D^{\prime}}.\vec{\tau}_l}\vec{u}^{D^{\prime}}_B(\vec{r})
\right]\nonumber\\
\end{eqnarray} 
\end{widetext}

with $t^{(1)}=t^{(2)}=t$, $t^{(3)}=t^{\prime}=t(1-2\epsilon)$, 
$\vec{k}^{D^{\prime}}.\vec{\tau}_3=0$, and $\vec{k}^{D^{\prime}}.\vec{\tau}_1=-\vec{k}^{D^{\prime}}.\vec{\tau}_2=-\theta$, where $\theta=\arccos\left(-\frac {t^{\prime}}{2t}\right)=a\frac{\sqrt{3}}2 k^D_x$.\

$h^{\prime AB}_{int}$ takes then the following form:
\begin{eqnarray}
h^{\prime AB}_{int}&=&-i \frac {3a}2\frac 1 b \frac{\partial t}{\partial b}
\left[
\left(1+\frac 23 \epsilon \right) u^{D^{\prime}}_{Ax}(\vec{r})
+i u^{D^{\prime}}_{Ay}(\vec{r})\right.\nonumber\\
&+&\left.\left(1-\frac 43 \epsilon \right) u^{D^{\prime}}_{Bx}(\vec{r})
-i\left(1+2 \epsilon \right) u^{D^{\prime}}_{By}(\vec{r})
\right]
\end{eqnarray}
where we considered the limit of small strain amplitude ($|\epsilon|\ll 1$).\

To bring out the signature of the electronic dispersion on the EPC, we assume that the phonon dispersion at $K^{\prime}$ point is not affected by the strain. 
This means that the phonon polarization of the highest frequency optical mode is \cite{Ando}
$\vec{e}_{D^{\prime}}=\vec{e}_{K^{\prime}}=\frac 12 =\left(
\begin{array}{c}
1\\
-i\\
1\\
i
\end{array}
\right)$.\

The matrix element $h^{\prime AB}_{int}$ becomes:
\begin{eqnarray}
h^{\prime AB}_{int}&=&-3ai\frac 1 a \frac{\partial t}{\partial a}
\left(1+\frac 13 \epsilon \right) \vec{e}_{D^{\prime}}^{\;0}.\vec{U}_{D^{\prime}}\nonumber\\
\end{eqnarray}
where 
\begin{eqnarray}
\vec{U}_{D^{\prime}}=\left(
\begin{array}{c}
\vec{u}^{D^{\prime}}_A\\
\vec{u}^{D^{\prime}}_B
\end{array}
\right)=\sqrt{\frac{\hbar}{2NM\omega_K}}\vec{e}_{K^{\prime}}\sum_{\vec{q}}
\left(b_{D,\vec{q}}+b^{\dagger}_{D,-\vec{q}}\right)\mathrm{e}^{i\vec{q}.\vec{r}}\nonumber\\
\end{eqnarray}
and $\vec{e}_{D^{\prime}}^{\;0}=\frac 1{2\left(1+\frac 13 \epsilon\right)}\left(
\begin{array}{c}
1+\frac 23 \epsilon\\
-i\\
1-\frac 43 \epsilon\\
i(1+2\epsilon)
\end{array}
\right)$.\

The effective interaction Hamiltonian takes a form similar to that found by Suzzura and Ando \cite{Ando}:
\begin{eqnarray}
H_{int}=-3\frac t {a} \beta_K \left(1+\frac 13 \epsilon\right)
\left(
\begin{array}{cc}
0 & \Delta_{D^{\prime}} \sigma_y\\
\Delta_{D} \sigma_y &  0
\end{array}
\right)
\end{eqnarray}
with $\Delta_{D^{\prime}}=\vec{e}_{D^{\prime}}^{\;0}.\vec{U}_{D^{\prime}}$, 
$\beta_K=-\frac b t \frac{\partial t}{\partial b}$ and $\sigma_y$ is the Pauli matrix.

\vspace{1cm}
$^{\ast}$ Electronic address: sonia.haddad@fst.rnu.tn

\end{document}